\documentclass[fleqn,usenatbib]{mnras}

\usepackage{newtxtext,newtxmath}

\usepackage[T1]{fontenc}

\DeclareRobustCommand{\VAN}[3]{#2}
\let\VANthebibliography\thebibliography
\def\thebibliography{\DeclareRobustCommand{\VAN}[3]{##3}\VANthebibliography}
\usepackage{xcolor}

\usepackage{graphicx}	
\usepackage{amsmath}	
\usepackage{amssymb}	
\usepackage{tablefootnote}
\usepackage{caption}

\title[Inflation of Migrated Hot Jupiters]{Inflation of Migrated Hot Jupiters}
\author[M. Mol Lous and Y. Miguel]{M. Mol Lous$^{1}$\thanks{E-mail:
mollous@strw.leidenuniv.nl} and Y. Miguel$^{1}$\thanks{E-mail: ymiguel@strw.leiden.nl}\\
$^{1}$Leiden Observatory, University of Leiden, Niels Bohrweg 2, 2333  CA Leiden, The Netherlands}

\pubyear{2020}

\begin{document}
\label{firstpage}
\pagerange{\pageref{firstpage}--\pageref{lastpage}}
\maketitle

\begin{abstract}
The observed low densities of gas giant planets with a high equilibrium temperature (hot Jupiters) can be simulated in models when a fraction of the surface radiation is deposited deeper in the interior. Meanwhile migration theories suggest that hot Jupiters formed further away from their host-star and migrated inward.
We incorporate disk migration in simulations of the evolving interior of hot Jupiters to determine whether migration has a long lasting effect on  the inflation of planets.
We quantify the difference between the radius of a migrated planet and the radius of a planet that formed in situ as the \textit{radius discrepancy}. We remain agnostic about the physical mechanism behind interior heating, but assume it scales with the received stellar flux by a certain fraction.\\
We find that the change in irradiation received from the host-star while the planet is migrating can affect the inflation and final radius of the planet. Models with a high fraction of energy deposited in the interior ( > 5\%) show a significant radius discrepancy when the deposit is at higher pressures than $P=1 \,bar$. For a smaller fraction of 1\%, there is no radius discrepancy for any deposit depth.
We show that a uniform heating mechanism can cause different rates of inflation, depending on the migration history. If the forthcoming observations on mean densities and atmospheres of gas giants give a better indication of a potential heating mechanism, this could help to constrain the prior migration of such planets.
\end{abstract}



\section{Introduction}
\begin{figure}
	\includegraphics[width=\linewidth]{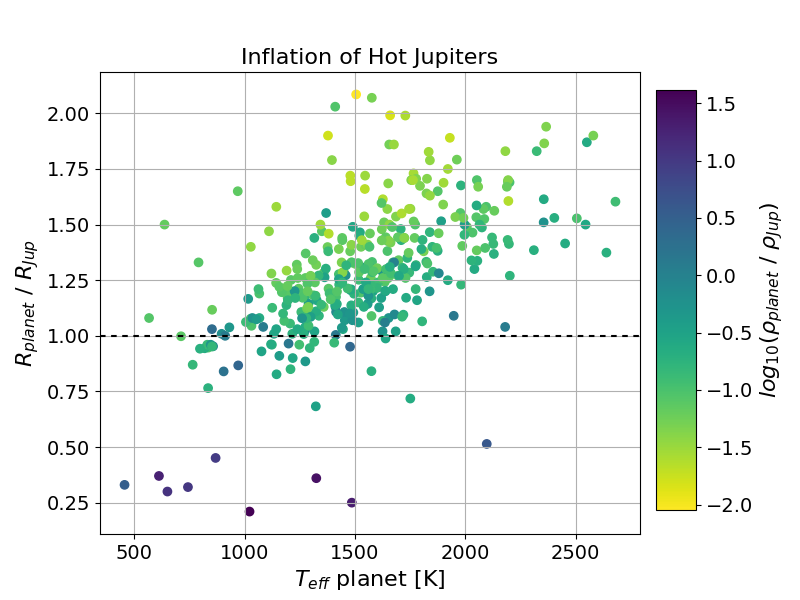}
	\captionsetup{width=.8\linewidth}
    \caption{Exoplanets with $0.3 \, M_{Jup} \, < \, M \, < \, 10 \, M_{Jup}$ and a period shorter than 10 days.} The data is obtained from  \href{http://exoplanet.eu/catalog/}{http://exoplanet.eu/catalog/}. Most planets that fit the definition of hot Jupiter have a density significantly below the density of Jupiter. The radius of these planets increases for their effective temperature.
    \label{fig:scatter_inflation}
\end{figure}

In the last 25 years we have seen the hot Jupiter grow from one anomalous discovery into a distinct population of detected exoplanets. In this timespan they have taken up an important role in exoplanet science.
Because they are massive planets at short periods, they are relatively easier to detect by radial velocity and transit photometry, making this a big population. They further encompass a majority of the planetary mass in a system and they can interact strongly with other objects in the system.
To achieve a better understanding of planetary systems, it is therefore necessary to understand the origin, formation and evolution of these planets.\\
One unexplained observed feature of hot Jupiters is their inflation. Their measured density is lower than simulations can reproduce, even when surface radiation from the host star is considered.
Simulations of hot Jupiters start with a high-entropy, inflated planet. During the evolution this inflation is expected to recede as the planet looses it's formation heat by cooling and contracting. To some extend the application of surface radiation can slow down the contraction, but this is insufficient to reproduce the observed inflation \citep{Baraffe_2003}.
Therefore, in order to reproduce inflation of hot Jupiters, alternative or additional physical mechanisms need to be considered.\\
There is an empirical trend between the equilibrium temperature of the planet and the rate of inflation.
For planets with equilibrium temperature above $1200 \, K$, the radius anomaly  (difference between the observed radius and the one that results from evolution models) increases as a powerlaw $R \propto T_{eq}^{1.4}$, while planets with an equilibrium temperature below $1000 \, K$ are deflated \citep{Laughlin_2011, Miller_2011, Demory_2011, Laughlin_2015}.When simulating the rate of inflation of warm, hot and ultra-hot jupiters with a known mass and radius, there is a relation between the planet's effective temperature and the necessary rate of incident flux converted to interior heat. This rate peaks (2.5\%) at an effective temperature of $T=1500$ while for gas giants with a higher or lower effective temperature this rate decreases \citep{Thorngren_2018}. Figure \ref{fig:scatter_inflation} shows the decrease in density for an increasing equilibrium temperature.
Due to this relation a potential heating mechanism is likely based on an interaction with the host star. 
\\
There are physical arguments for such heating mechanisms. 
One of them is \textit{Hydrodynamical dissipation}. According to this theory heat gets transported into the interior of the planet. Such a mechanism could be based on vertical winds pushing down kinetic energy that is subsequently dissipated into heat \citep{Guillot_2002, Showman_2002}, heat moving downward by turbulent mixing in the radiative zone \citep{Youdin_2010} or a vertical advection of potential temperature \citep{Tremblin_2017, Sainsbury_2019}. The second radiation dependent mechanism is \textit{Ohmic dissipation}, in which electric currents move from the atmosphere to the interior \citep{Batygin_2010, Perna_2010, Huang_2012, Rauscher_2013, Wu_2013, Rogers_2014, Ginzburg_2016}. These currents could theoretically be present in planets that have an equilibrium temperature of $T_{eq} > 1200 \, K$.
Either of these mechanisms can be simulated in interior models by taking a fraction of the received stellar radiation and depositing it deeper in the interior.
\\
Previous studies have reproduced the observed inflation in models of giant planets when incorporating this extra interior heating.
Without specifying a certain mechanism behind the extra heat deposit, but basing it on the stellar irradiation, \citet{Baraffe_2003} found that when $1 \%$ of the received flux is transferred to internal heat in the planets layer that contains $2 \cdot 10^{-5}$ of it's total mass, the inflation of \textit{HD 209458b} can be reproduced. \citet{Komacek_2017} showed that transferring the same fraction to a depth of $10^{2} \, bars$ suffices for the case of \textit{HD 209458b}. Furthermore they showed that heating in deeper layers increases inflation up to a depth of $10^{4} \, bars$. A deeper deposit than $P_{dep}=10^{4} \, bars$ does not increase inflation. A similar study by \citet{Ginzburg_2015} deposited extra radiation at different optical depths to induce an extra convective layer and radiative layer, which pushed the RCB deeper in the interior, delaying cooling and resulting in enough inflation to match observations.\\
These models were under the assumption that the planet received a constant amount of stellar radiation throughout the evolution. More recent simulations by \citet{Komacek_2020} applied the same interior heating deposit at planets which were already contracted, causing reinflation.
These simulations showed that reinflation within the host-stars main-sequence timescale requires deeper deposits than was the case for inflation due to inhibited cooling.\\
The hot jupiter reinflation of \citet{Komacek_2020} was based on an increase of stellar radiation as it evolves on the main-sequence, and applied on completely cooled planets. These simulations assumed that the planets did not change their semimajor axis with respect to the star, but rather that they formed in situ. \\
In this work, the authors did not consider the change in irradiation from the star or evolution of the planet prior to the main sequence stage. In this paper, we evaluate how the migration history and change in stellar irradiation during planet formation might affect the planet inflation. Formation theories of hot Jupiters favour the planet being created at a distance further away from the host star and migrating inward after the formation. \textit{Disk migration} specifically involves the planet moving in along with the dust and gas of the protoplanetary disk during the accreting phase of the star. This implies a change in received radiation that could have a significant and long lasting effect on the evolution of the planet.
Since the ratio of inflation is scattered for a given equilibrium temperature (see figure \ref{fig:scatter_inflation}), this scatter could be caused by different migration histories of the planets.\\
The aim of this study is to determine if the migration of hot Jupiters early in their evolution can have a lasting effect on their observed inflation. This paper will commence with the method in section \ref{sec:method}. Here the interior model will be discussed and how the interior heating and migration are applied. The results are presented and discussed in section \ref{sec:results}. In the conclusion, section \ref{sec:conclusion}, the results are summarized. This is followed by a brief outlook for potential improvements for future research.

\section{Method} \label{sec:method}
\subsection{Interior Model}
The hot Jupiter interior evolution is simulated by using the code \textit{MESA}. This code creates a one-dimensional stellar interior model and evolves it by solving the stellar structure equations. A \textit{MESA} model of a gas giant planet underwent a few alternations from the stellar variant. The first two used equations describe the mechanical structure of the planet:\\
\begin{equation} \label{eq:fss_1}
   \frac{\mathrm{d} m}{\mathrm{d} r} = 4 \pi r^{2} \rho(r)
\end{equation}
\begin{equation} \label{eq:fss_2}
    \frac{\partial P}{\partial m} = - \frac{G m }{4 \pi r^{4}}
\end{equation}\\
Equation \ref{eq:fss_1} relates a shell with infinitesimal thickness $dr$ and mass $dm$ to the density $\rho (r)$ at a radial distance $r$. Equation \ref{eq:fss_2} uses hydrostatic equilibrium to relate the pressure $P$ in this shell to the enclosed mass $m$, with $G$ the gravitational constant.\\
\begin{equation}
    \frac{\partial l}{\partial m} = \epsilon_{grav} + \epsilon_{extra}
    \label{eq:fss_3}
\end{equation}
\begin{equation}
     \frac{\partial T}{\partial m} = - \frac{GmT}{4 \pi r^{4} P } \nabla
     \label{eq:fss_4}
\end{equation}\\
Evolution of the model is calculated by energy conservation (equation \ref{eq:fss_3}) and energy transport (equation \ref{eq:fss_4}). Shell $dm$ radiates energy $dl$ based on the local change in gravitational energy $\epsilon_{grav}$ and extra interior energy $\epsilon_{extra}$.  This extra energy is based on received stellar radiation and deposited in the interior, following \citet{Komacek_2017}. A star with nuclear burning would have an extra term for the change in local energy, but this is not applicable in the planetary models. In the equations $T$ is the temperature and $\nabla$ is the logarithmic temperature gradient ($d \, ln P / d \, ln T$) , that could be either radiative or adiabatic, following the Schwarzschild criterion. To solve this set of equations an equation-of-state is used that is designed for hydrogen and helium in giant planets and low mass stars \citep{Saumon_1995}. The implemented opacities are from \citet{Freedman_2008}. \\
Based on the core accretion model for giant planet formation, the model starts by creating a $10\, M_{\earth}$ core of heavy material. Subsequently an envelope of H/He is added, resulting in a model with a total mass of $1 \, M_{Jup}$ and radius of $2 \, R_{Jup}$. We assume that the planet accretes all it's mass before the start of migration. \\
The initial radius, $2 \, R_{Jup}$ in our simulations, would not be relevant for a solely cooling model. A model with a larger initial radius and thus a higher entropy model, would have a higher luminosity in the early stage of the evolution and thus cool down more rapidly. However, since the migration is initiated before such models converge, the initial radius might determine at what evolution the planet start to receive the radiation, which could influence the outcome.

\subsection{Classic type II Migration}
We simulate the planet to undergo \textit{classic type II migration}. During formation, when a planet accretes a certain critical mass, it can open up a gap in the protoplanetary disk. This is the expected case for a Jupiter-sized planet \citep{Crida_2006, Baruteau_2014}. The consequence is that the planet exerts a larger torque on the disk than the viscous torque of the disk itself. The disk is then split up: an inner- and outer disk are formed, with a planet in between. The torque of the inner disk pushes the planet outwards, while the outer disk pushes it in, leaving the planet stuck in the gap. During the accreting phase of the star, both gas and planet will move to closer orbits. \\
In this situation the migrating planet will have a radial speed that is identical to the gas accretion speed, which is \citep{Shakura_1973}:\\
\begin{equation} \label{eq:accr_speed}
\left (\frac{da}{dt} \right )_{visc} = - \frac{3}{2} \alpha_{\nu} h^{2} \sqrt{G M_{\star} /a}
\end{equation}\\
This speed depends on the stellar mass $M_{\star}$, the semimajor axis $a$ and parameters of the protoplanetary disk. The unitless parameter $\alpha_{\nu}$ characterizes the viscosity. The parameter $h$ is the scale height of the disk, defined as approximately the sound speed divided by the Kepler frequency, locally in the disk. By integrating equation \ref{eq:accr_speed}, the timescale for a planet to migrate from an initial semi-major axis $a_{0}$ to a final semi-major axis $a_{end}$ is:\\
\begin{equation}
    \label{eq:semimajor_axis_time}
    \tau_{migr} = \frac{ 4 \left( a_{0}^{3/2} - a_{end}^{3/2}  \right ) } {9\alpha_{\nu} h^{2} \sqrt{G M_{\star}}}
\end{equation}\\
Using equation \ref{eq:semimajor_axis_time} we can estimate the timescale of migration for a Jupiter-sized planet around a solar-mass star, migrating from $10 \, AU$ to $0.05 \, AU$. For this we assume a common value of $\alpha_{\nu}=0.003$ and take into account that h is between $\sim0.09$ and $\sim0.02$ for a semi major axis between 10 and 0.05 AU. Approximating $h$ as $0.05$ we obtain a timescale of around $\sim 3 \cdot 10^{5}\, years$\\
The actual timescale of migration, however, can be expected to deviate from this significantly. The radial variation of disk properties is not simple and transitions in temperature and surface density profiles, among other factors, can affect the migration timescales \citep{Cridland_2016}. In addition, the disk parameters $\alpha_{\nu}$ and $h$ can have a range of values that result in different timescales. The estimated $\alpha_{\nu}$ values are broadly distributed between orders of magnitude ($1e{-4} $ to $ 1e{-2}$) and there is not yet a correlation found with global properties of the disk or star \citep{Rafikov_2017, Ansdell_2018}. The value does seem correlated with the disk's rate of ionization, which implies that $\alpha_{\nu}$ is not a constant but changes during the evolution of the disk \citep{Martin_2019}. This would induce an extra time variability in the accretion speed of equation \ref{eq:accr_speed}. Another reason to expect deviating timescales of migration is that \textit{classic type II migration} probably assumes a too ideal case in which the gap opened up by a planet is completely clear of gas and dust. Hydrodynamical simulations show that a planet can create a gap with a much lower density, but still a fraction of gas remains passing this gap. This would result in a decoupled planetary migration that can differ from the viscous accretion speed \citep{Duffel_2014, Durmann_2015}. Simulations can even have a planet migrating through a stationary disk \citep{Robert_2018}. A more realistic representation of migration speeds might be the empirical results of such simulations \citep{Kanagawa_2018, Ida_2018}.\\
Since a more detailed description of planetary migration is out of the scope of this paper, we explore the possibility of different timescales by introducing an extra variable C, that will be referred as to the delay constant.
The migration speed and timescale from equation \ref{eq:accr_speed} including this constant becomes:\\
\begin{equation}
    \frac{da(t)}{dt} = - \frac{3}{2} \, C \, \alpha_{\nu} h^{2} \sqrt{G M_{\star} /a}
\end{equation}\\
The semi-major axis at a given time $t > t_{0}$ will then be:\\
\begin{equation}
    a(t) = \left ( a_{0}^{\frac{3}{2}} - C \frac{9}{4} \alpha_{\nu} h^{2} \sqrt{G \, M_{\star}}  (t - t_{0}) \right )^{\frac{2}{3}}
    \label{eq:migr_delay}
\end{equation}\\
With a corresponding migration timescale of,
\begin{equation} \label{eq:timescale_with_delay}
    \tau_{migr}=\frac{1}{C} \frac{ 4 \left( a_{0}^{3/2} - a_{end}^{3/2}  \right ) } {9\alpha_{\nu} h^{2} \sqrt{G M_{\star}}}
\end{equation}\\
Once $a(t)=a_{end}$ is reached the semi-major axis remains constant.\\
For inward migration $C$ needs to be a positive number. Another constraint is that the disk migration needs to be completed before the disappearance of the gas disk, which can take up to 10 million years \citep{Haisch_2001, Ribas_2015}. This implies a lower limit of $C \approx 0.05$.

\subsection{Applying received flux as a variable}
\label{subsec:apply_frec}
\begin{figure}
	\includegraphics[width=1\linewidth]{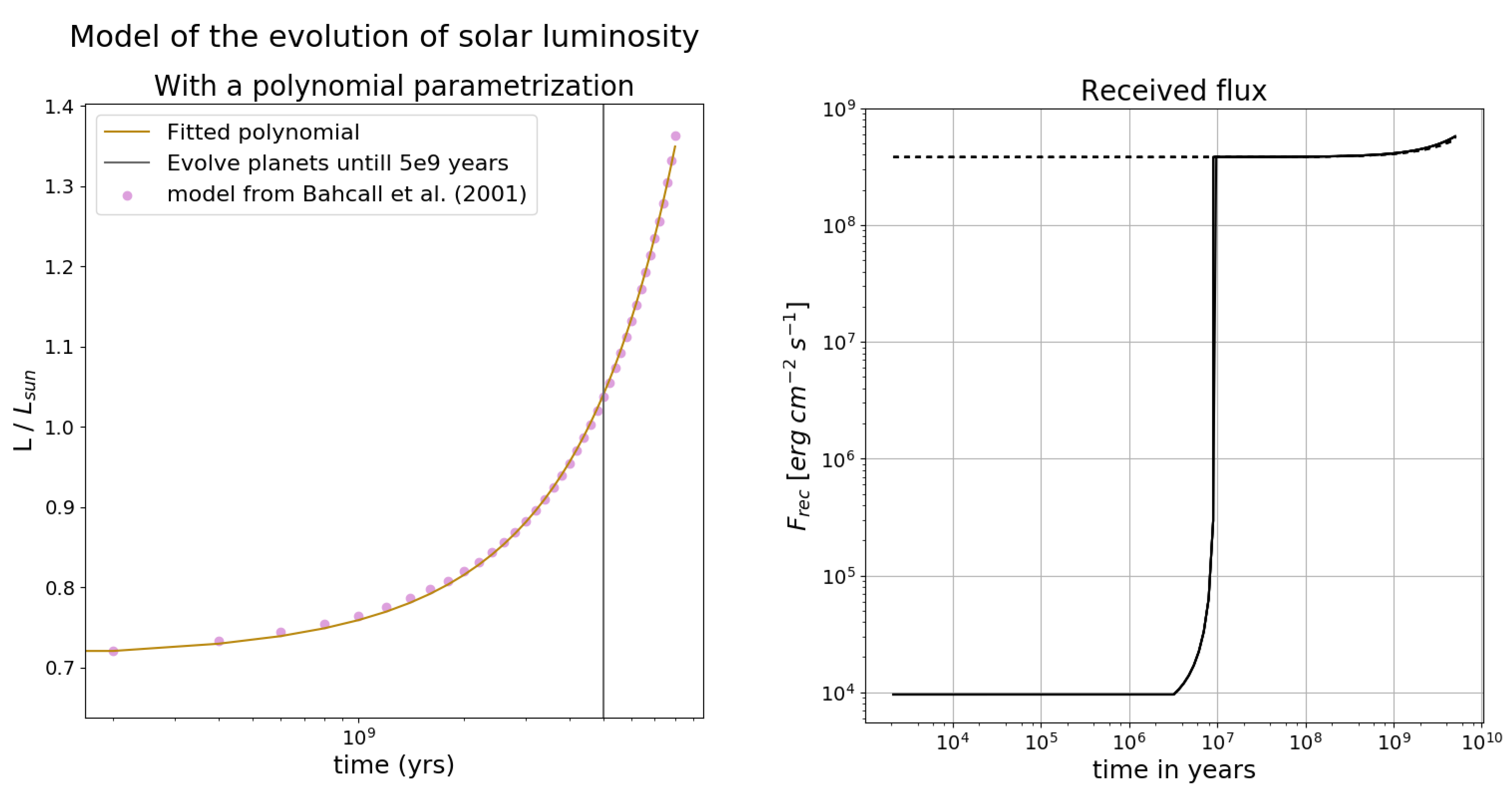}
    \caption{Left: Evolution of bolometric luminosity by \citet{Bahcall_2001}. Planets are evolved to an age of 5 billion years, when the luminosity is that of the current sun. Right: the received flux applied in the models based on \textit{Classic migration} as detailed in table \ref{tab:migration} (solid line) compared to in situ formation (dashed line). The effect of the stars luminosity evolution is negligible compared to the change in flux due to migration.}
    \label{fig:luminosity}
\end{figure}

In addition to the changing semi-major axis, the received flux is varied based on an evolving stellar luminosity. We simulate the luminosity using a model of the sun by \citet{Bahcall_2001} as it evolves on the Main Sequence. This luminosity evolution is shown in figure \ref{fig:luminosity}. The star has an almost constant luminosity of $ \sim 0.72 \, L_{\odot}$ during the first $ \sim 10^{8}$ years. Luminosity evolution of the star has therefore little influence on the evolution of the planet during the migration. It will become relevant after disk migration is finished, as is shown in section \ref{sec:results}.
 \\
We use two methods to insert extra energy in the model based on the calculated received radiation. The first is \textit{surface radiation}. Applying this in \textit{MESA} requires the assignment of two variables: received flux $F_{rec}$ and column-depth $\Sigma_{p}$. The surface radiation gets uniformly distributed over the outer layers of the model where $\Sigma \leq \Sigma_{p}$. The extra energy per unit mass $\epsilon_{irr}$ then is:\\
\begin{equation} \label{eq:surface_irr}
    \epsilon_{irr} = \frac{F_{rec}}{4 \Sigma_{p}}
\end{equation} \\
Values from $150$ to $300\; g / cm^{2}$ are in agreement with commonly used opacities in the visible \citep{Fortney_2008, Guillot_2010, Owen_2016}. In these simulations $\Sigma_{p} = 220 \, g / cm^{2}$ was used, which corresponds to an opacity of $\kappa_{\nu}=0.005 \, cm^{2} \, g^{-1}$.
In addition to surface radiation, \textit{extra interior heat} is applied deeper in the model. The subroutine used to insert this extra heat was taken from \textit{MESA} Summerschool material\footnote{Made by Jonathan Fortney in 2017 and available at: \href{http://cococubed.asu.edu/mesa_market/education.html}{cococubed.asu.edu/mesa\_market/education.html}}. An interior heating mechanism is assumed for which a fraction of the received radiation gets converted to extra heat: $E_{exta} = F_{rec} \cdot \pi R_{pl}^{2}\cdot f_{int}$, with  $f_{int} << 1$. This fraction gets deposited around a chosen depth in the model as a Gaussian distribution. The Gaussian peaks at the \textit{depth of deposit}, $P_{dep}$, similar to prior simulations by \citet{Komacek_2017}. The interior heating is much smaller than the surface radiation in terms of energy. However, due to the deep deposit it can have a significant effect on the cooling efficiency of the planet and consequently on the rate of inflation.\\

\subsection{Chosen Parameters}
Table \ref{tab:migration} summarizes the default values of \textit{classic disk migration}.
\begin{figure}
	\includegraphics[width=\linewidth]{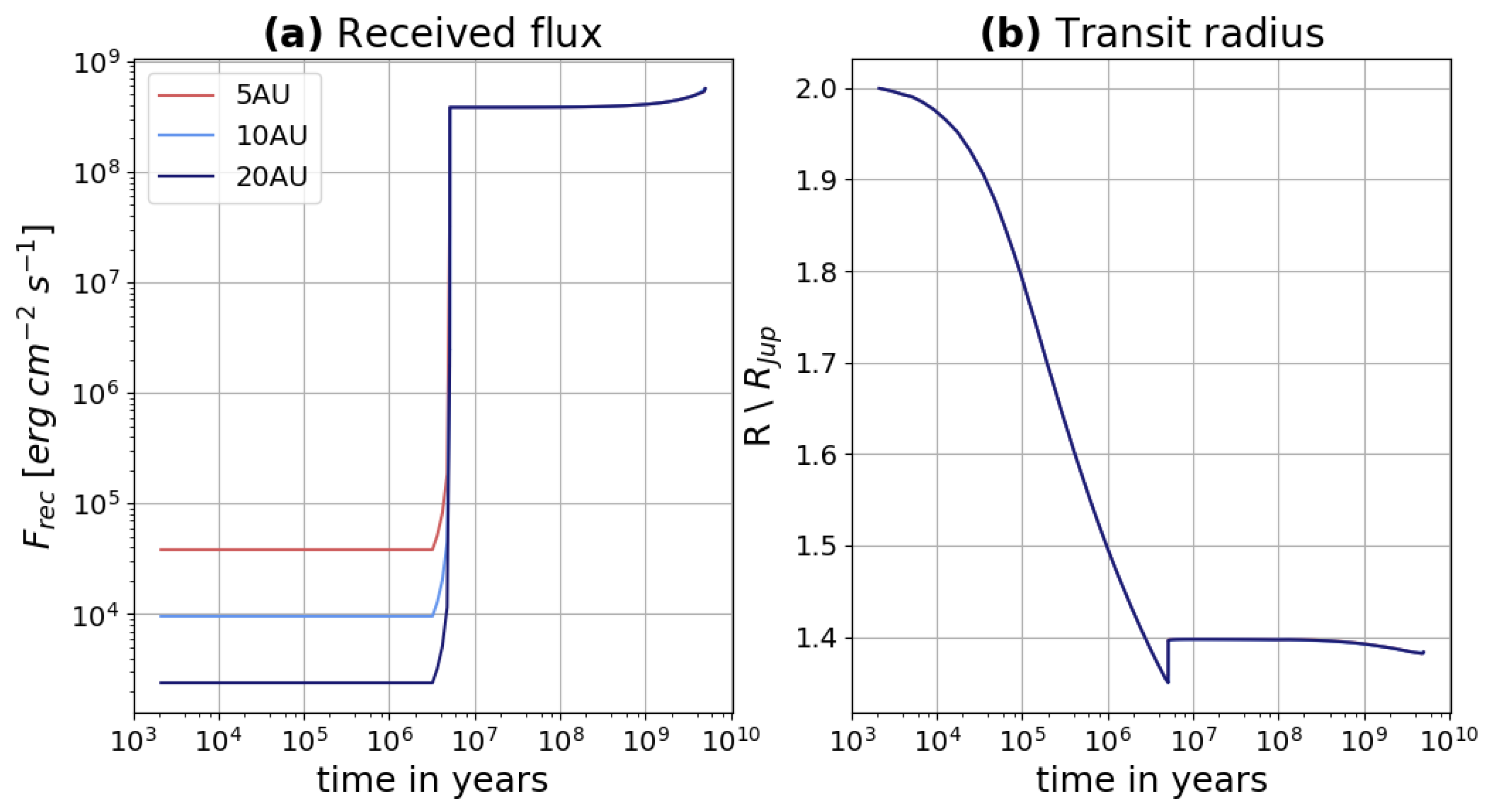}
	\captionsetup{width=.9\textwidth}
    \caption{Migrating planet with internal heating at depth $P_{dep}=10^{3}$ and fractions of extra internal heat $f_{int}=0.1$. Migrations starting from 5, 10 or 20 AU. The delay constant $C$ in equation \ref{eq:timescale_with_delay} is adjusted so that the planet arrives at $a=0.05 \, AU$ after 1e7 years. \textbf{(a):} The received flux during the evolution. \textbf{(b):} The radius evolution, which is the same for all migration paths.}
    \label{fig:vary_start}
\end{figure}
\begin{figure*}
    \captionsetup{width=.8\textwidth}
	\includegraphics[width=\textwidth]{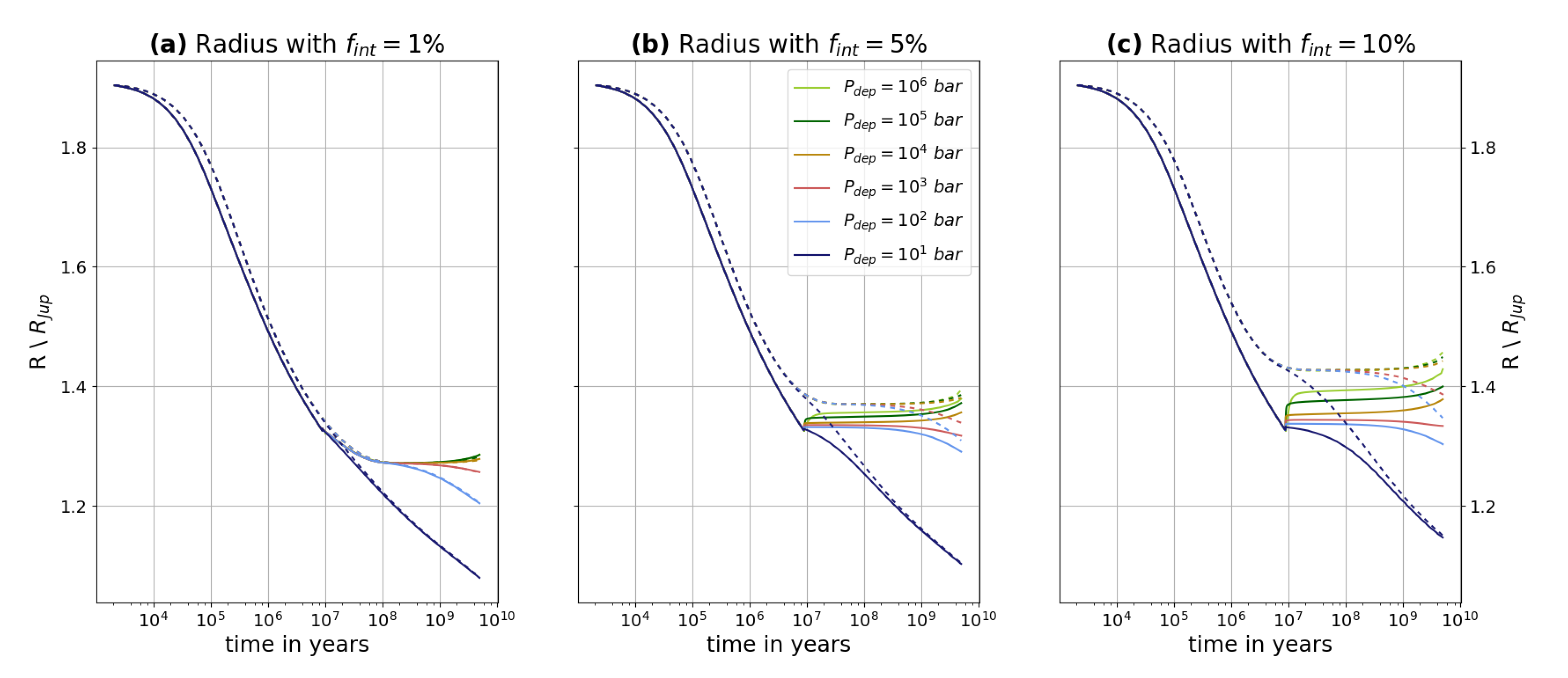}
    \caption[width=.8\textwidth] {The radius evolution of models when 1\% (a), 5\% (b) or 10\% (c) of the received flux is deposited at different pressure-depths in the interior. A higher fraction of interior heating increases the eventual radius. When 5\% and 10\% are inserted deeper than 10 bar there is a discrepancy between migrated and constantly radiated models.}
    \label{fig:radii}
\end{figure*}
Migration starts after the formation of the planet, which will be the zero-age in our simulations. \\
Because observations on protoplanetary disks show that these disks last between 1 and 10 million of years \citep{Ribas_2015}, we pick the timescale of migration, $\tau_{migr}$, as 7 million years, to also take into account the time for forming the planet . We assume this formation time is $t_{0}=$ 3 million years \citep{Alibert_2018} and the end of migration is at $t=1e7$ years. To match this timescale the delay constant is chosen as $C=0.05$ in these calculations.\\
The migration will stop at $0.05 \, AU$ since this is the orbit at which hot jupiters are relatively often found, even when a detection bias is subtracted \citep{Heller_2018}. The equilibrium temperature that corresponds to this distance, assuming a sun-like star is $1254 \, K$.\\
The planet is formed at $10 \, AU$, which is chosen to have formation outside of the snowline where \textit{core accretion} is more likely due to a large feeding zone \citep{Dawson_2018}. Hot Jupiter formation theory implies that they formed around $\sim 5 \; AU$. In our calculations the location of formation has little effect on the evolution, since we constrain the arrival time at $t=1e7$ years by adjusting the delay constant. This makes the $a_{0}$ and $C$ two degenerate variables. Figure \ref{fig:vary_start} shows simulations where formation starts at $5, \, 10, \, $ and$ \, 20 \, AU$. In order to have migration end at $t=1e7$ years, the delay constant is chosen as $C= 0.05,\,0.14, \, 0.4 $.\\
The values $f_{int}=1\%$ and $P_{dep}=10^{3}\, bar$ were found by \citet{Komacek_2017} as sufficient to reproduce inflation.
\begin{table} 
\centering
\begin{tabular}{l|l|l|l}
\multicolumn{1}{l|}{}             & Classic Model               & Migr parameter & Source \\ \hline
\multicolumn{1}{l|}{$a_{0}$}      & 10 AU                               & $\checkmark$        &    \citet{Dawson_2018}     \\
\multicolumn{1}{l|}{$a_{end}$}    & 0.05 AU                         & $\checkmark$        &   \cite{Heller_2018}     \\
\multicolumn{1}{l|}{$t_{0}$}    & 3e6 yrs                         & $\checkmark$        &     \citet{Alibert_2018}    \\
\multicolumn{1}{l|}{C}           & 0.05 $^a$                        & $\checkmark$        &     \citet{Ribas_2015}    \\
\multicolumn{1}{l|}{$\Sigma_{p}$} & $220 \; g \; cm^{-2}$  & $\times$            &   \citet{Paxton_2013}      \\
\multicolumn{1}{l|}{$f_{int}$}    & 1\%                              & $\times$            &     \citet{Komacek_2017}    \\
\multicolumn{1}{l|}{$P_{dep}$}    & $10^{3} \; bar$         & $\times$            &     \citet{Komacek_2017}    \\ \hline
\multicolumn{4}{l}{$^a$ This delay constant is chosen to end migration at $t=1e7\, yrs$.}\\ 
\end{tabular}
\caption{The default parameters used in the simulations. If other values are used this will be specified. Migration parameters are those parameters that determine the semi-major axis of the planet over time. Non-migration parameters determine the application of radiation, which depends on the received flux.} \label{tab:migration}
\end{table}
After the migration is finished, we continue the evolution of the planet to an age of 5 Gyr, which is a typical age of a main sequence star. Once this age is reached, the radius is calculated at which the optical depth $\tau \approx 1$, approximately the radius measureable by transit measurements.\\

\section{Results and Discussion} \label{sec:results}

\begin{figure}
    \centering
	\includegraphics[width=0.9\linewidth]{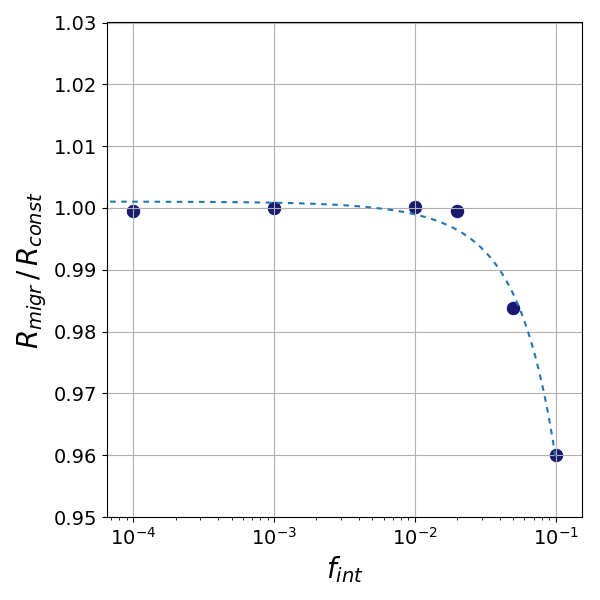}
	\captionsetup{width=.9\textwidth}
    \caption{Radius discrepancy of hot Jupiter models that migrated and those that evolved in situ: $R_{migr} / R_{const}$. The depth of internal heat deposit is fixed at $P_{dep}=10^{3} \, bar$ and fractions of extra internal heat vary. For low fractions $f_{int}$, there is no discrepancy while there is some for fractions above 1 \%.}
    \label{fig:ratio_frac}
\end{figure}

\begin{figure*}
	\includegraphics[width=0.8\textwidth]{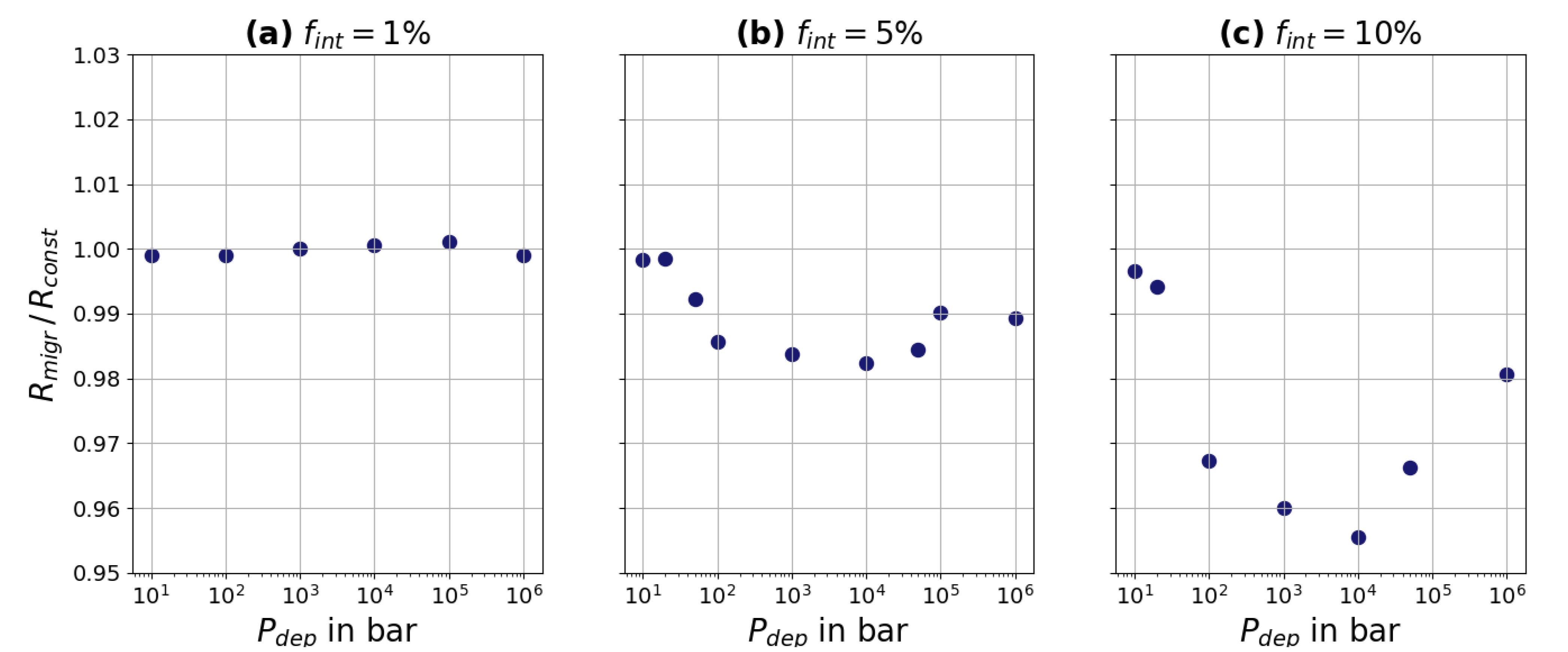}
    \caption[width=.8\textwidth]{Radius discrepancy of hot Jupiter models that migrated and those that evolved in situ: $R_{migr} / R_{const}$. The fraction is fixed while the depths of deposit vary. For a fraction of 1\%, none of the depths cause a discrepancy. For a fraction of 5\% or 10\%, a deeper injection increases the discrepancy up to a depth of $\sim 10^{4} \; bar$. For injections at higher pressures, the discrepancy decreases slightly.}
    \label{fig:ratio_dep}
\end{figure*}
\begin{figure}
	\includegraphics[width=\linewidth]{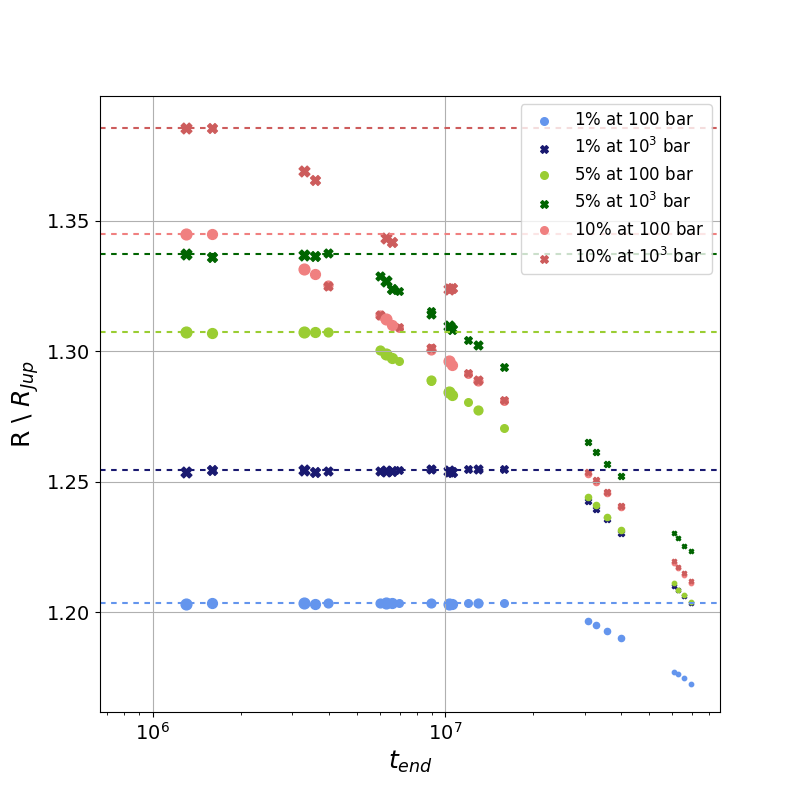}
	\captionsetup{width=.8\textwidth}
    \caption{The radii of migrated hot Jupiter models after evolving to 5 Gyrs. $t_{end}$ $=t_{0} + \tau_{migr}$ indicates when the planet reached the inner radius of $0.05 \, AU$. Different colors indicate different parameters (fraction, depth) for the extra interior heating. Dashed lines are the radii of models that evolved in situ.}
    \label{fig:arrival_time}
\end{figure}

In this paper we take into account the change in the irradiation received by the planet due to two effects: 1) the evolution of the stellar luminosity with time and 2) the change in semimajor axis as a consequence of planetary migration. Figure \ref{fig:luminosity} (right) shows that both effects are relevant at a different timescales: up to 10 Myr planetary migration has a large impact in the received flux, while the increase of stellar luminosity affects the long term evolution up to the Gyrs.\\
Figure \ref{fig:radii}, $\textbf{(a, b, c)}$, show the evolution of the transit radii of classically migrating models (solid lines) compared to models that evolve in situ (dashed lines). The extra interior energy is applied as 1\% (a), 5\% (b) and 10\% (c). For each of these fractions, the energy is injected at a depths of $10$, $100$, $10^{3}$, $10^{4}$, $10^{5}$ or $10^{6}$ $bar$.\\
The plots show that larger energy deposits result in larger radii. It is furthermore shown that a deeper deposit results in a larger radius discrepancy up to a depth of $10^{4} \, bar$. This is in line with previous results by \citet{Komacek_2017}, and due to the depth of the energy determining the size of the radiative zone of the planet. The size of this zone can constrain the efficiency of cooling and contraction \citep{Fortney_2010, Thorngren_2019}\\ Figure \ref{fig:radii} shows that planets that migrated have a smaller final radius than those that were formed in situ. This difference in radii can be expressed as the ratio Rmig/Rconst, which we call the radius discrepancy. This is because those formed in situ received high irradiation during their entire lives, while those that migrated received lower irradiation at the beginning of their formation history. Only after migrating they were more exposed to high irradiation from the central star.\\
A difference in eventual radius between a migrated and a constantly irradiated model is present for higher fractions of 5\% and 10\% and when the depth is $100$ bars or deeper. For a lower fraction of 1\% the radii of both models are the same after 5 Gyr. For this interior heating mechanism migration does not influence the eventual observed radii.\\
Figure \ref{fig:ratio_frac} and \ref{fig:ratio_dep} show the radius discrepancy (Rmig/Rconst) for different models. As seen in figure \ref{fig:ratio_frac}, when the depth is constant at $P_{dep}=10^{3} \, bar$, there is a growing radius discrepancy for fractions above 1\%. In figure \ref{fig:ratio_dep} the fraction of deposited energy is kept constant while the depths are varied. For a constant deposited of 1\% (a), the ratio remains around one, an expected result considering figure \ref{fig:radii}. The fractions 5\% (b) and 10\% (c) have a growing radius discrepancy for deeper deposits, up to a maximum around $10^{4}\, bar$. As a consequence of the re-inflation of planets exposed at a varying luminosity with time, deposits at higher pressures result in a smaller discrepancy. This figure shows \textit{that in the case of an interior heating mechanism which is strongly related to the incident flux, migration will influence the degree of inflation.}
\begin{figure*}
	\includegraphics[width=\textwidth]{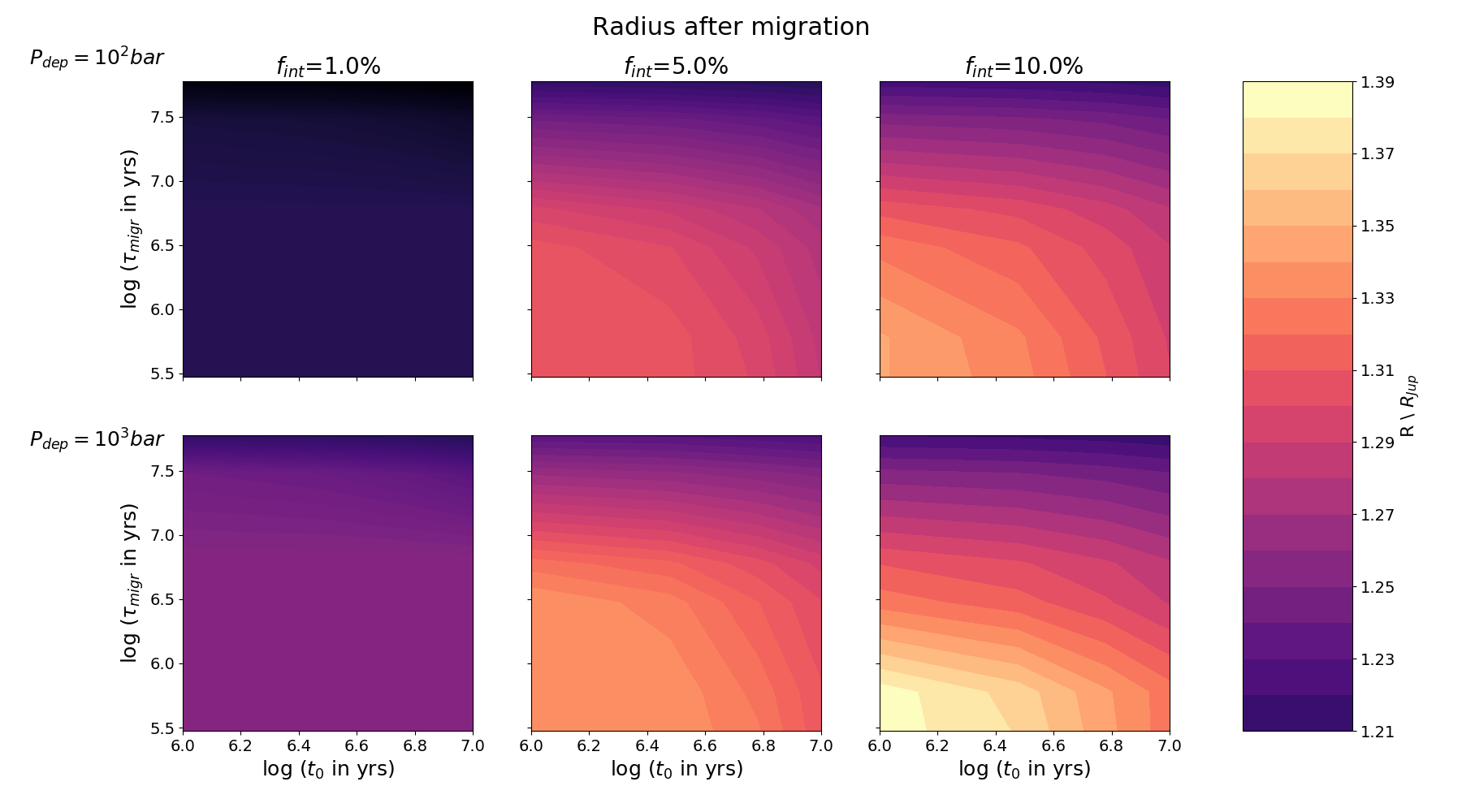}
	\captionsetup{width=.8\textwidth}
    \caption{The radii of evolved hot Jupiter models, varying the fraction ($f_{int}=0.1, \, 0.5, \, 0.1$) of received flux that is applied as interior heating and the depth ($P_{dep}=10^{2} \, , 10^{3}\, bar$) at which it is deposited. When a small fraction is applied, migration does not influence the rate on inflation. For energy deposits of 5\% or 10\%, the eventual radius does depend on the start of migration $t_{0}$ (x-axis) and the duration of migration $\tau_{migr}$ (y-axis).}
    \label{fig:migr_param}
\end{figure*}
Up to now a certain migration is assumed, where different interior heating parameters in combination with this migration determine the eventual radius. Alternatively, there can be a fixed interior heating, so that different migrations will lead to different radii. For six different combinations of $f_{int}$ and $P_{dep}$, the radii of models after different migrations are shown in figure \ref{fig:migr_param}. The  migration can vary in timescale $\tau_{migr}$ or starting time $t_{0}$. The timescale is varied by using different constants $C$ in equation \ref{eq:migr_delay}.\\
Once again there is no discrepancy for a 1\% injection. The uniform color demonstrates that any combination of $\tau_{migr}$ and $t_{0}$ leads to the same radius. The higher fractions of 5\% and 10\% already proved to lead to a discrepancy in combination with a depth of $100$ or $10^{3}$ $bar$. Figure \ref{fig:migr_param} shows that this discrepancy depends on the start and duration of migration as well. A fraction of 10\% leads to significantly a higher radius than 5\% when the timescale of migration is short. Changing the depth from $100$ $bar$ to $10^{3}\; bar$ also leads to larger radii in the case of fast migration. Migration with longer timescales lead to the same radii, independent of the fraction or depth of the extra heat.\\
Since the received flux scales with $\propto a^{-2}$, the time spend at close-in orbits could be more significant for the evolution than the start of migration. Considering this, and the fact that the speed of migration increases as the semi-major axis decreases, a more characteristic parameter of migration could be the time it reaches the final orbit $a_{end}$. This happens at $t_{end} = t_{0} + \tau_{migr}$.\\
Figure \ref{fig:arrival_time} shows the radii of evolved planets for different $t_{end}$. The value of $t_{end}$ in this figure range from 1 Myr to 60 Myr. While the figure shows that longer arrival times result in less inflation, it should be noted that the longer arrival times would be long after the disappearance of the protoplanetary disk at $\sim10$ Myr \citep{Ribas_2015}. Alternative migration mechanisms, such as eccentricity scattering or tidal dissipation, could work on longer timescales as they are applicable after the disk has disappeared.\\
The dashed lines indicate the radius of models that evolved in situ. For migrated models that have the radius on this line the inflation is not related to the migration history. This is the case for the fraction of 1\%, at least for migrations within the disks lifetime.
A relation between radius and arrival at $a_{end}$ starts to show at arrival timescales beyond disk migration.
Larger energy deposits give a steeper relationship between arrival time and radius. This shows once again that for more injected energy, the eventual radius is more sensitive to the migration.\\
The final radii from 1\% deposits follow a line with little dispersion. This indicates that the inflation is closely related to the time the planet arrives at the final orbit, but not to the duration of migration. For larger deposits, there is a scatter in radius for a certain $t_{end}$. In this case inflation does depend on the combination of start and duration of migration.
\begin{figure}
	\includegraphics[width=\linewidth]{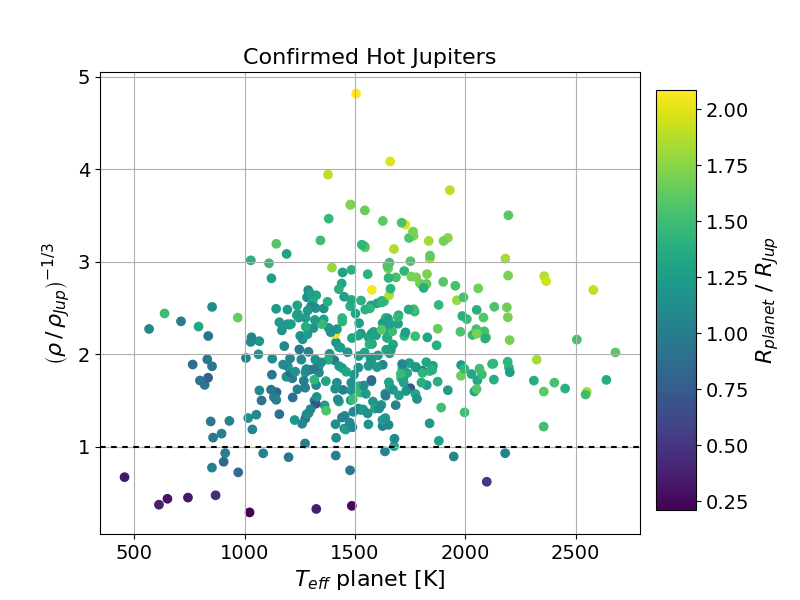}
	\captionsetup{width=.9\textwidth}
    \caption{Exoplanets with $0.3 \, M_{Jup} \, < \, M \, < \, 10 \, M_{Jup}$ and a period shorter than 10 days, obtained from \href{http://exoplanet.eu/catalog/}{http://exoplanet.eu/catalog/}.}
    \label{fig:scatter_inf_density}
\end{figure}\\
To compare the different eventual radius of models with different migration histories to the observed scatter in inflation we use figure \ref{fig:scatter_inf_density}. Like figure \ref{fig:scatter_inflation}, it shows the observed hot jupiters. The inflation is shown using the density as $\left ( \rho \, / \, \rho_{Jup} \right ) ^{-1/3} = \left ( M \, / \, M_{Jup} \right ) ^{-1/3} \cdot \left ( R \, / \, R_{Jup}\right )$. Since all the simulated models assume a constant mass of $1 \, M_{Jup}$, the scatter on the y-axis of figures \ref{fig:arrival_time} and \ref{fig:scatter_inf_density} can be compared.\\
The observed scatter in hot jupiter inflation is much bigger than the different radii resulting from the simulations. The models in figure \ref{fig:arrival_time} have an equilibrium temperature of $1254 \, K$. Assuming that the arrival time of type II migration can range between 2 to 10 million years and applying an interior heating based on $10 \, \%$ of the incident flux can result in a radius difference of $0.05 \, R_{Jup}$. In the case of $5 \, \%$, this difference can be $0.02 \, R_{Jup}$. Meanwhile the observed densities at $T_{eq}=1254 \, K$ in figure \ref{fig:scatter_inf_density} corresponds to a much bigger spread in radii. Therefore, while migration is a relevant mechanism that affects the inflation history of the planets, a variation in disk migration arrival times alone is not sufficient to reproduce the scatter in observed inflation.\\
\\
The conditions under which migration can have a lasting effect on the observed inflation of hot jupiters favour a high fraction of stellar flux being deposited in the interior. This fraction is somewhat higher than what is estimated for the theoretical interior heating mechanisms that depends on the incident stellar flux. The interior heating parameters of Ohmic dissipation are estimated as up to $f_{int }\lesssim 5 \%$ \citep{Perna_2010, Rauscher_2013}. In terms of depth this mechanisms will loose most of the heat in the atmosphere. Only a small fraction of the deposited energy will reach the interior, where it can significantly influence the radius evolution \citep{Rauscher_2013}. Tidal dissipation can result in a higher fraction of $f_{int} \approx 10 \%$ \citep{Arras_2010}, but can also be on the order of $0.1 \%$ \citep{Bodenheimer_2001}.

\section{Conclusion} \label{sec:conclusion}
In this paper we performed simulations of planetary evolution to study the influence of inward migration on the observed planetary radius. Our models explore different migration timescales, take a stellar flux into account that varies with time and consider different inflation mechanisms by adding extra energy at different pressures in the planet interior. Our results show that migration can influence the observed radius of the planet in the case of a strong, stellar flux dependent, interior heating.\\
The inflation of hot Jupiters could be a reflection of the amount of radiation received at earlier stages in the planet's formation. Particularly the time at which the planet reaches the inner orbit is related to the rate of inflation. Planets that arrive at their observed close orbital distances at a later stage can be significantly less inflated, even though they are subjected to the same heating mechanism. However, we find that while migration most likely played a role in the final inflation history of the planets, it is not the main and only cause of the scatter in inflation.\\
This relation furthermore requires that the deposited energy is based on a large fraction ($\gtrsim  5\%$) of the incident flux. For smaller fractions of energy deposit, the migrating planet evolves to the same radius as a planet that evolves under constant, high radiation. Furthermore the depth of the deposit needs to be beyond $10$ bar. We found an evident relation between the amount of extra energy and the discrepancy in radius between migrated and non-migrated models.\\
Our simulations migrated planets to a distance of $0.05 \, AU$ around a sun-like star. This corresponds to an equilibrium temperature of $T_{eq}=1245 \, K$. The radius discrepancy will probably be larger for planets that reach a higher equilibrium temperature, as the different migration arrival times result in a larger difference in received flux. We did not consider migration timescales beyond a few $\sim \, 10$ Myrs, since those are beyond the estimated timescales of disk migration. Larger timescales could, however, be the result of alternative mechanisms which can migrate the planet after the disk has disappeared. High-eccentricity tidal migration can work on planets when they are perturbed into a high-eccentricity orbit, by mechanisms such as planet-planet scattering \citep{Rasio_1996, Weidenschilling_1996, Chatterjee_2008} or the Kozai-Lidov mechanism \citep{Kozai_1962, Lidov_1962, Naoz_2016}, followed by a tidal migration inwards.\\
\\
The simulated interior heating is not based on a specific physical mechanism and therefore there is a lot of freedom in the parameters. A better understanding of the heating mechanisms that acts on inflation inflation would constrain these parameters. We would have a better understanding of the depth at which the energy is deposited and of whether this deposit is related to the stellar flux. This would lead to a better understanding of the influence of planet migration and its imprint in the planetary radius. \\
Current and future observations could improve this understanding. TESS is expected to add many warm jupiters to the discovered sample, which might help understand the relation between effective temperature and inflation \citep{Grunblatt_2019}.
The recently launched CHEOPS will accurately measure the radii of sub-Neptunes and super-Earths of which the mass is already know from radial-velocity measurements \citep{Broeg_2013}. This should result in a better mass-radius relationship of such planets. A statistical analysis of Neptune-sized planets would show if they are inflated to any extend. PLATO will aim at improving the mass-radius relationship of hot Jupiters in combination with stellar parameters. This will broaden the potential statistical relationships with inflation and could lead to a better understanding of the nature of this phenomenon \citep{Rauer_2014}. Finally ARIEL, planned to launch in 2028, will measure the atmospheric composition of known exoplanets. More information on the atmospheric composition could also lead to a better insight in potential heating mechanisms.

\section*{Acknowledgements}
The authors thank the anonymous referee for the insightful comments which improved the paper. We furthermore thank J. Fortney for providing the subroutine to include internal heating in the planet's interior. We thank the Exoplanet Group at Leiden Observatory for useful discussions that help improving this paper. 




\bibliographystyle{mnras}
\bibliography{mnras_template} 








\bsp	
\label{lastpage}
\end{document}